\documentclass[twocolumn,showpacs,preprintnumbers]{revtex4-2}
\usepackage[caption=false]{subfig}
\usepackage{amsmath}  
\usepackage{amsfonts}  
\usepackage{graphicx} 
\usepackage{enumerate}
\usepackage{tikz}
\usepackage{xcolor}
\usetikzlibrary{shapes,arrows, arrows.meta, calc, positioning}
\usepackage[colorlinks=true,linkcolor=blue,urlcolor=blue,citecolor=blue]{hyperref}
\begin{document}
\newcommand{\ds}{\displaystyle}
\newcommand{\be}{\begin{equation}}
\newcommand{\en}{\end{equation}}
\newcommand{\bea}{\begin{eqnarray}}
\newcommand{\ena}{\end{eqnarray}}

\title{Could a thin-shell configuration lie hidden within the Universe?}
\author{Mauricio Cataldo}
\altaffiliation{mcataldo@ubiobio.cl} 
\affiliation{Departamento de
F\'\i sica, Universidad del Bío-B\'io, Casilla 5-C,
Concepci\'on, Chile.\\}
\affiliation{Centro de Ciencias Exactas, Universidad del Bío-Bío, Casilla 447, Chillán, Chile.}
\author{Antonella Cid}
\altaffiliation{acidm@ubiobio.cl} 
\affiliation{Departamento de
F\'\i sica, Universidad del Bío-B\'io, Casilla 5-C,
Concepci\'on, Chile.\\}
\affiliation{Centro de Ciencias Exactas, Universidad del Bío-Bío, Casilla 447, Chillán, Chile.}
\author{Pedro Labraña}
\altaffiliation{plabrana@ubiobio.cl} 
\affiliation{Departamento de
F\'\i sica, Universidad del Bío-B\'io, Casilla 5-C,
Concepci\'on, Chile.\\}
\affiliation{Centro de Ciencias Exactas, Universidad del Bío-Bío, Casilla 447, Chillán, Chile.}

\date{\today} 
\begin{abstract}
This article explores the cosmological scenario in which our Universe contains a hidden thin-shell configuration. We investigate a degenerate modification of the Friedmann-Robertson-Walker metric obtained through a coordinate transformation applied to the radial coordinate, analogous to recent approaches that address the Big Bang singularity via spacetime defects. The resulting metric, while formally satisfying the standard homogeneous Friedmann equations, actually describes an evolving wormhole geometry with two asymptotically flat Friedmann-Robertson-Walker regions connected by a throat located at the coordinate singularity. Using Israel's junction formalism, we demonstrate that this coordinate singularity corresponds to a thin shell characterized by exotic matter with well-defined surface energy density and isotropic pressure. The shell obeys the barotropic equation of state $p = -\rho/2$, confirming the presence of exotic matter that violates the standard energy condition, which is a requirement for maintaining wormhole geometries. As the universe expands, this thin shell becomes increasingly diluted, scaling as $1/a(t)$ with the cosmic scale factor.

\vspace{0.5cm}
\end{abstract}
\smallskip
\maketitle 


\section{Introduction}
Modern cosmology takes as its starting point the Friedmann solution to Einstein's gravitational field equations, assuming matter is modeled as a homogeneous perfect fluid. The standard flat Friedmann-Robertson-Walker (FRW) cosmology is written as
\begin{eqnarray}
ds^2= -dt^2+a^2(t) \left(dr^2+r^2(d \theta^2+\sin^2\theta \, d\phi^2) \right), \label{FRWmetric}
\end{eqnarray}
where $a(t)$ is the scale factor. For a homogeneous perfect fluid with energy density 
$\rho$ and isotropic pressure $p$, the Einstein equations are given by
\begin{eqnarray}
\kappa \rho= 3\left(\frac{\dot{a}}{a}\right)^2, \label{EFRW1} \\
\kappa p=-2\frac{\ddot{a}}{a}-\left(\frac{\dot{a}}{a}\right)^2, \label{EFRW2}
\end{eqnarray}
where $\kappa=8 \pi G$ and the derivatives with respect to cosmic time $t$ are expressed using an overdot.
To complete this set of equations, one must specify an equation of state, that is, a relation between the pressure and energy density of the perfect fluid, $p=p(\rho)$. Additionally, the matter content is assumed to satisfy the standard energy conditions.

It is widely recognized that the cosmological solutions to these equations suggest an early point in time when the energy density and curvature of the universe were infinitely high; this initial moment is known as the Big Bang. 

Since a singularity signals a breakdown of the laws of physics, several efforts have been made in the literature to avoid the Big Bang scenario. For example, the authors in Refs.~\cite{Malkiewicz:2009qv,B0} construct a model that yields a regular cosmological solution and analyze both the underlying mechanism and its main features. In this framework, the initial singularity is replaced by a cosmological bounce. This transition from a Big Bang to a Big Bounce, within the context of an FRW universe, is achieved through classical-level modifications of the theory inspired by loop quantum geometry~\cite{Malkiewicz:2009qv} and the underlying structure of loop cosmology~\cite{B0}. See also Refs.~\cite{B1,B2,B3} for other approaches capable of resolving the Big Bang singularity via a bounce. 

On the other hand, ekpyrotic cosmologies~\cite{E1,E2,E3,E4} represent a class of theoretical models in which the universe does not originate from a singularity, but rather from a slow contraction phase, followed by a bounce that leads to the expanding universe we observe today.

Alternatively, there are emergent scenarios in which the universe is assumed to originate from an Einstein static state, subsequently evolving into an inflationary era. In these models, the Big-Bang singularity is avoided, as the universe emerges from a past-eternal static configuration~\cite{Ellis:2002we,Ellis:2003qz,2007mp,Labrana:2018bkw,PL}.

Recently, the concept of a spacetime defect has been suggested in the literature as a way to address the Big Bang singularity~\cite{K,KPRD1,KPRD2,KPRD3,KPRD4}. This type of object can be represented by a degenerate metric, which has a vanishing determinant on a three-dimensional submanifold of spacetime, while a nonzero length scale is introduced to smooth out the Friedmann Big Bang singularity. This modified metric is referred to as the regularized-Big Bang metric~\cite{Battista}.

The author of Refs.~\cite{K,KPRD1,KPRD2} demonstrates that a regularized version of the FRW metric~(\ref{FRWmetric}) can be achieved by introducing a time-dependent function into the metric component $g_{tt}$, which modifies the Friedmann equations and smooths out the curvature singularity at $t=0$. We review this procedure in the following section.

This paper is organized as follows: In Sec.~II, we briefly outline the key aspects of the regularized FRW Big Bang singularity. Section~III focuses on a defect wormhole embedded in a flat FRW universe. In Sec.~IV, we analyze the thin-shell structure within the flat FRW universe. Finally, our conclusions are presented in Sec.~V.

\section{Regularized FRW big bang singularity}\label{SectionII}
The elimination of the Big Bang curvature singularity in the Friedmann cosmological solution has been investigated as a fundamental challenge in modern cosmology. To address this issue, the authors of Ref.~\cite{K} propose a modification of the FRW metric through the following degenerate spacetime metric ansatz:
\begin{eqnarray}
ds^2=-\frac{t^2}{t^2+b^2} \, dt^2+a(t)^2 \left(dr^2+r^2(d \theta^2+\sin^2\theta \, d\phi^2) \right), \nonumber \\ \label{modifiedmetric}
\end{eqnarray}
where $t \in (-\infty, \infty)$ and the length scale $b$ appearing in the metric component $g_{tt}$ is conventionally taken to be positive. The standard FRW metric~(\ref{FRWmetric}) is recovered when $b \to 0$.

The crucial feature of the modified metric~(\ref{modifiedmetric}) is that, for $b > 0$, it becomes degenerate at $t = 0$, where $g_{tt} = 0$ and $\det(g_{\mu\nu}) = 0$. Physically, this corresponds to a three-dimensional spacetime defect located on the $t = 0$ hypersurface.

While there exists a direct relation between the FRW metric~(\ref{FRWmetric}) and the ansatz metric~(\ref{modifiedmetric}), a crucial distinction emerges upon closer examination. Consider the coordinate transformation from $t \in (-\infty, \infty)$ to a new time coordinate $\tau$ defined by~\cite{K}:
\begin{equation}
\tau = \begin{cases}
+\sqrt{b^2 + t^2}, & \text{for } t \geq 0, \\
-\sqrt{b^2 + t^2}, & \text{for } t \leq 0. \label{TRF}
\end{cases}
\end{equation}
Under this transformation, the modified metric~(\ref{modifiedmetric}) takes the standard form of the FRW metric~(\ref{FRWmetric}) in terms of $\tau$:
\begin{equation}
ds^2= -d\tau^2+a^2(\tau) \left(dr^2+r^2(d \theta^2+\sin^2\theta \, d\phi^2) \right),
\label{modifiedmetricR}
\end{equation}
where the cosmic time $\tau \in (-\infty, -b] \cup [b, \infty)$.

A key property of the coordinate transformation~(\ref{TRF}) is its multivalued nature at $t = 0$, which results in a discontinuous inverse transformation. Since diffeomorphisms require invertible $C^\infty$ mappings, this transformation cannot be classified as such. The absence of diffeomorphic equivalence indicates that the differential structures of the metrics~(\ref{FRWmetric}) and~(\ref{modifiedmetric}) are genuinely distinct, despite their formal similarity.
 
In summary, the metrics~(\ref{FRWmetric}) and~(\ref{modifiedmetric}) are metrically equivalent but differentially distinct. This fundamental difference implies that they do not simply represent alternative coordinate systems for the same spacetime, but rather describe physically different spacetimes.

\section{Defect Wormhole Embedded in a Flat FRW Universe}
We now apply a similar procedure to the FRW metric~(\ref{FRWmetric}), but with an analogous transformation applied to the radial coordinate $r$, thus obtaining a degenerate form of the FRW metric. We begin by introducing a coordinate transformation given by 
\begin{eqnarray} 
r^2(\xi) = \lambda^2 + \xi^2, \label{Txi} 
\end{eqnarray} 
which is analogous to Eq.~(\ref{TRF}), but applied to the radial coordinate $r$ instead. Here $\lambda$ is assumed to be a positive real parameter and $-\infty<\xi<\infty$. Specifically, the function $r(\xi)$ in Eq.~(\ref{Txi}) is not one-to-one within any neighbourhood of $\xi = 0$ that includes both positive and negative values of $\xi$.

Therefore, from Eqs.~(\ref{FRWmetric}) and~(\ref{Txi}), one obtains the line element
\begin{eqnarray}
ds^2= -dt^2+a^2(t) \left(\frac{\xi^2 d\xi^2}{\lambda^2+\xi^2}+(\lambda^2+\xi^2) \, d^2\Omega \right),  \label{defectFRWmetric}
\end{eqnarray}
where $d^2\Omega=d \theta^2+\sin^2\theta \, d\phi^2$ is the standard line element on the unit 2-sphere. 

Based on the form of metric~(\ref{defectFRWmetric}), one might expect that the dynamical equations would include inhomogeneous terms depending on $\xi$ and its derivatives. However, it can be shown that the Einstein equations in this case are the standard homogeneous Friedmann equations~(\ref{EFRW1}) and~(\ref{EFRW2}), which determine both the scale factor and the cosmic matter content.

Therefore, one might ask: what does the metric in question represent? We note that for the metric in Eq.~(\ref{defectFRWmetric}), its covariant form $g_{ab}$ is smooth ($C^{\infty}$) throughout the entire manifold. However, the $g_{\xi \xi}$ component in Eq.~(\ref{defectFRWmetric}) becomes zero at $\xi=0$; therefore, the inverse metric $g^{ab}$ is not continuous at this point. Its singular character is further reinforced by the fact that at $\xi=0$ the metric determinant also vanishes, which causes the metric to degenerate at that point (note that the metric determinant also vanishes when $a(t)=0$).

The metric~(\ref{defectFRWmetric}) has another important characteristic. The angular section of this metric is $(\lambda^2+\xi^2) \, a^2(t) \, d^2\Omega$, which represents a two-dimensional sphere of radius $\sqrt{\lambda^2+\xi^2} \, a(t)$. For $\xi=0$, the angular section of the metric becomes $\lambda^2 \, a^2(t) \, d^2\Omega$, which represents a two-dimensional sphere of radius $\lambda \, a(t)$. In this way, at $\xi=0$ the two-dimensional spheres reach their minimum radius $\lambda \, a(t)$. This behavior is typical of wormhole configurations; therefore, we conclude that the metric~(\ref{defectFRWmetric}) admits the interpretation of an evolving wormhole: for each constant time $t=t_0$, we have $a(t_0)=\text{constant}$, and the spacetime~(\ref{defectFRWmetric}) corresponds to a Lorentzian wormhole geometry with two asymptotically flat regions — one as $\xi \to \infty$ and the other as $\xi \to -\infty$. The spherical throat is located at $\xi = 0$, with a radius of $a(t_0)\lambda$ and an area of $4\pi a^2(t_0)\lambda^2$. Accordingly, the line element~(\ref{defectFRWmetric}) characterizes a zero-tidal-force flat wormhole spacetime whose evolution is dictated by the scale factor $a(t)$. 

In this context, a natural question arises: if the dynamical Einstein equations of the spacetime~(\ref{defectFRWmetric}) are given by the homogeneous Friedmann equations, what kind of anisotropic matter supports this evolving wormhole (or at least its throat)? 

To address this issue more clearly, we consider evolving wormholes that exhibit similar dynamical behavior. There exists a family of evolving wormholes whose expansion is also determined by the scale factor that satisfies the Friedmann equations~(\ref{EFRW1}) and~(\ref{EFRW2}). This family of wormholes is sustained by two matter components: one with homogeneous energy density $\rho(t)$ and isotropic pressure $p(t)$, and another with inhomogeneous energy density $\rho_{in}(t, r)$, and anisotropic pressures $p_r(t,r)$ and $p_t(t,r)$. Remarkably, the matter content that threads and sustains such wormholes is the inhomogeneous and anisotropic component, while the rate of expansion of the evolving wormhole is determined by the homogeneous and isotropic component~\cite{CataldoEWH}. In this context, the field equations for such wormholes are typically inhomogeneous due to the presence of matter threading and sustaining them (see Eqs.~(3)–(6) of Ref.~\cite{CataldoEWH}). From this observation, we conclude that the evolving wormhole described by Eq.~(\ref{defectFRWmetric}) is not supported by any anisotropic and inhomogeneous matter content.

At first glance, one might conclude that the geometry~(\ref{defectFRWmetric}) still describes a flat FRW universe, expressed in an unusual coordinate system, with its expansion governed by the scale factor \( a(t) \) and the standard Friedmann equations. However, a more careful analysis reveals that this interpretation is misleading. In fact, the spacetime~(\ref{defectFRWmetric}) constitutes a degenerate form of the metric in Eq.~(\ref{FRWmetric}), and it actually describes a time-evolving wormhole geometry, in which two regions of the flat FRW universe are connected through a throat located at $\xi=0$.

However, it would be premature to conclude that the absence of an inhomogeneous and anisotropic matter component necessarily implies that the wormhole throat satisfies the standard energy conditions. Such a conclusion would contradict what we know about wormholes, which must be supported by exotic matter that violates the energy conditions at the throat.

The degeneration of the FRW metric at $\xi=0$ thus reveals the presence of a wormhole structure hidden by the pathological choice of coordinates. We demonstrate that this throat constitutes a thin shell. In this configuration, each asymptotic region maintains its own cosmological evolution governed by the Friedmann equations with scale factor $a(t)$, but the connection between the two regions is made through a surface of discontinuity where exotic matter is concentrated that violates the energy conditions. Therefore, the metric does not represent a single FRW universe in unusual coordinates, but rather the dynamic union of two flat FRW cosmologies through a wormhole structure whose throat evolves simultaneously with the expansion of the connected regions.

In summary, the Einstein equations derived for the metric~(\ref{defectFRWmetric}) are homogeneous because this degenerate spacetime—despite its seemingly complex form that allows for the description of a wormhole due to its explicit dependence on the radial coordinate $\xi$—actually corresponds to a highly symmetric geometry. It remains homogeneous and isotropic on each spatial slice of constant cosmic time, with the sole exception of a thin-shell layer at $\xi=0$, as we shall see in the next section.

\section{Thin-shell within flat FRW Universe}\label{SectionIV}
As stated above, the wormhole throat is located at $\xi = 0$. At this location, the metric coefficient $g_{\xi\xi}$ in Eq.~(\ref{defectFRWmetric}) vanishes, while the angular components remain finite. This distinctive behavior suggests a coordinate singularity at $\xi = 0$. However, it may also indicate the presence of a thin shell located at this position. We now show that the vanishing of $g_{\xi\xi}$ at $\xi = 0$ indeed implies the presence of such a shell. To establish this, we must analyze the extrinsic curvature and examine its continuity (or lack thereof) across the hypersurface. A discontinuity in the extrinsic curvature, according to Israel's junction formalism, corresponds to a non-vanishing surface energy-momentum tensor. This represents a physical distribution of matter-energy localized on the hypersurface $\xi=0$, which constitutes the precise definition of a thin shell in general relativity.

To obtain the characteristics of the matter that sustains the thin shell at $\xi=0$, we begin by calculating the extrinsic curvature of hypersurfaces $\xi=constant$. To accomplish this, we consider the unit normal vector to these hypersurfaces in the form
\begin{eqnarray}
N_a=\left(0,\frac{a(t) |\xi|}{\sqrt{\lambda^2+\xi^2}},0,0 \right), \\
N^a=\left(0,\frac{\sqrt{\lambda^2+\xi^2}}{a(t) |\xi|},0,0 \right).
\end{eqnarray}
It is straightforward to verify that $N^a \cdot \partial_t=0$, $N^a \cdot \partial_\theta=0$ and $N^a \cdot \partial_\phi=0$. Note also that this vector points in the direction of increasing $\xi$.

In the coordinate system~(\ref{defectFRWmetric}), the extrinsic curvature of the constant hypersurfaces $\xi=constant$ can be calculated using~\cite{Poisson} 
\begin{equation}
K_{ab} = N_{(a;b)} = \frac{1}{2} \mathcal{L}_N g_{ab},
\label{eq:extrinsic_curvature}
\end{equation}
where $\mathcal{L}_N$ is the Lie derivative along the normal vector $N$.

The non-vanishing covariant components of the extrinsic curvature take the form
\begin{align}
K_{\theta\theta} &= a(t) \operatorname{sign}(\xi) \sqrt{\lambda^2+\xi^2},  \\
K_{\phi\phi} &= a(t) \operatorname{sign}(\xi) \sqrt{\lambda^2+\xi^2} \sin^2\theta,
\end{align}
from which we obtain
\begin{align}
K^\theta_\theta &=K^\phi_\phi=\frac{\operatorname{sign}(\xi)
}{a(t) \sqrt{\lambda^2+\xi^2}},
\end{align}
and the trace of the extrinsic curvature is given by
\begin{equation}
K = K^a_a = K^\theta_\theta + K^\phi_\phi = \frac{2 \,\operatorname{sign}(\xi)}{a(t)\sqrt{\lambda^2+\xi^2}}.
\end{equation}
From these expressions, we can calculate the jump in extrinsic curvature across the shell, which is given by
\begin{equation}
[K^a_b] = K^a_b(\xi=0^+) - K^a_b(\xi=0^-).
\end{equation}
Therefore:
\begin{align}
[K^\theta_\theta] &=[K^\phi_\phi]= \frac{2}{a(t)\lambda},
\end{align}
and the jump in the trace is
\begin{equation}
[K] = \frac{4}{a(t)\lambda}.
\end{equation}
From these expressions, we can compute the surface stress-energy tensor. According to Israel's junction conditions, it is given by 
\begin{equation} 
\kappa S^i_j = -\left([K^i_j] - \delta^i_j [K]\right).
\end{equation} 
This tensor is related to the surface energy density $\sigma$ and the surface isotropic pressure $\mathsf{P}$ through 
\begin{equation} 
S^i_j = \mathrm{diag}(-\sigma,\, \mathsf{P},\, \mathsf{P}). 
\end{equation}
Thus, we obtain for $S^a_b$:
\begin{align}
\kappa S^t_t &= \frac{4}{a(t)\lambda}, \\
\kappa S^\theta_\theta &= \kappa S^\phi_\phi = \frac{2}{a(t)\lambda}.
\end{align}
This surface stress-energy tensor corresponds to a thin shell with a surface energy density $\kappa \sigma = -\frac{4}{a(t)\lambda}$ and stresses in the angular directions: $\kappa \mathsf{P} = S^\theta_\theta = S^\phi_\phi = \frac{2}{a(t)\lambda}$. Therefore, the equation of state of this thin shell is
\begin{eqnarray}
\mathsf{P} = -\frac{\sigma}{2},
\label{SES}
\end{eqnarray}
which confirms the existence of a thin shell at $\xi=0$ with well-defined physical properties. 
It is worth noting that the matter supporting the thin shell possesses a negative energy density, indicating that it is entirely composed of exotic matter. In this case, we also find that $\mathsf{P} + \sigma = -\frac{2}{\kappa a(t) \lambda} < 0$, confirming that this exotic matter violates the Weak, Null, and Dominant energy conditions. However, it satisfies the Strong Energy Condition, as $\sigma + 2\mathsf{P} = 0$.

Clearly, as this geometry expands, the thin shell becomes diluted as $1/a(t)$.


\section{Conclusions}
In this work, we have investigated the possibility that our Universe could contain a hidden thin-shell configuration by applying a coordinate transformation to the radial coordinate of the standard FRW metric. This approach, analogous to recent attempts to address spacetime singularities through degenerate metrics, has revealed several significant findings.

Our primary result demonstrates that what initially appears to be merely an unusual coordinate representation of the flat FRW universe actually describes a fundamentally different spacetime geometry. The resulting degenerate metric characterizes an evolving wormhole with two asymptotically flat FRW regions connected through a throat located at $\xi = 0$. Remarkably, despite this complex geometric structure, the Einstein field equations reduce to the standard homogeneous Friedmann equations, indicating that the spacetime remains homogeneous and isotropic on each spatial slice of constant cosmic time, except at the location of the thin shell.

In this scenario, unlike conventional evolving wormholes that require anisotropic and inhomogeneous matter $\rho_{in}(t,\xi)$ distributed throughout space to sustain their geometric structure~\cite{CataldoEWH}, the wormhole described by the degenerate metric~(\ref{defectFRWmetric}) confines all exotic matter exclusively to a thin shell located at $\xi = 0$. Throughout the remaining spacetime, the anisotropic and inhomogeneous energy density is exactly zero, allowing the asymptotically flat regions to maintain the homogeneous  and isotropic cosmological evolution governed by the standard Friedmann equations.

Using Israel's junction formalism, we have established that the coordinate singularity at $\xi = 0$ corresponds to a thin shell characterized by exotic matter. The surface stress-energy tensor of this configuration exhibits a well-defined equation of state~(\ref{SES}), confirming the presence of exotic matter that violates the standard energy conditions.

The surface energy density scales as $\sigma \propto 1/a(t)$. As the universe expands, the thin shell becomes increasingly diluted while maintaining its fundamental exotic character, since the equation of state does not change in time. In the cosmological context, the surface energy density of this thin shell dilutes more slowly than ordinary matter ($\propto 1/a^3$) and radiation ($\propto 1/a^4$), but more rapidly than a cosmological constant. Therefore, this configuration could have been cosmologically significant in early epochs of the universe, when $a(t)$ was considerably smaller. Consequently, the continuous expansion of the universe makes these structures progressively more difficult to detect. Note that in the static limit of this evolving model, each slice with $a(t) = constant$ corresponds exactly to the so-called vacuum defect wormhole presented in Ref.~\cite{DW Klinkhamer}, which features a thin shell located at $\xi= 0$ whose equation of state is also $\mathsf{P} = -\sigma/2$~\cite{Visser}. 

This finding is consistent with our previous discussion, where we emphasized that this cosmological wormhole does not require  anisotropic and inhomogeneous matter distributed throughout space to sustain its geometry, since all exotic matter is concentrated exclusively within a thin shell located at $\xi=0$. The rest of the spacetime remains free of the exotic matter typically required to sustain such geometric structures.

This work requires an important clarification regarding its scope: while it reveals the potential existence of hidden wormhole structures within our universe, it does not resolve the Big Bang singularity, as the scale factor $a(t)$ continues to satisfy the standard Friedmann equations and the cosmological singularity at $t=0$ remains present. The main contribution lies in demonstrating that apparently homogeneous cosmological spacetimes may conceal complex spatial topologies supported by localized distributions of exotic matter, where the degeneration of the spacetime metric at the spatial hypersurface $\xi=0$ provides a distinctive diagnostic tool for identifying these hidden junction surfaces. From a broader perspective, these findings contribute to the ongoing exploration of alternatives to standard cosmological models by showing that such exotic matter configurations can be naturally embedded within FRW spacetimes without disrupting their overall dynamical evolution.

A natural extension of this work is to investigate the possible origin of the hidden thin-shell wormhole. Understanding the formation mechanism of such configurations remains an important open problem. Although such configurations cannot arise as classical solutions supported by ordinary matter, since their existence entails violations of the energy conditions, they could have been realized in the very early Universe. In this context, thin-shell wormholes could have emerged as topological defects during early-universe phase transitions, where quantum effects could have played an essential role in inducing wormhole geometries that would be classically prohibited. The exotic matter content characterized by $\mathsf{P} = -\sigma/2$ can naturally arise from quantum processes in curved spacetime, particularly through conformal anomaly–induced effective actions for matter fields typical of Grand Unified Theories. Within this same framework, it has been shown that such effective actions can even induce the formation of primordial wormholes in spherically symmetric configurations under specific initial conditions in the early Universe~\cite{Nojiri}. In another scenario, wormholes may form from supercritical domain walls nucleated during inflation via repulsive gravitational effects~\cite{Vilenkin,P2}.

Regardless of the specific formation mechanism, these configurations may encode imprints of pre-inflationary physics or quantum gravitational phenomena embedded in the classical geometry. While our analysis establishes the mathematical consistency and well-defined properties of this thin-shell configuration within classical general relativity, a complete understanding of its formation would require a dedicated study of thin-shell dynamics in the primordial Universe.

A further extension could involve multiple thin-shell wormholes distributed throughout an expanding FRW universe, following the statistical approach of Ref.~\cite{P2} for domain wall networks. In our case, such a network, with each structure diluting as $1/a(t)$, might originate from quantum topological fluctuations in the very early universe. However, a comprehensive analysis of multi-wormhole cosmologies would require the development of new mathematical tools to address the breakdown of perfect spatial homogeneity while preserving overall statistical isotropy.

Since the surface energy density scales as $\sigma \propto 1/a(t)$, this configuration would have been most relevant in the early universe, potentially leaving imprints on the CMB. 
In the current epoch, the thin shell would be highly diluted, making direct detection extraordinarily difficult. Although potential observational signatures could in principle emerge through gravitational lensing or gravitational-wave emission~\cite{Bambi,Shaikh}, detecting such exotic configurations would demand exceptionally large values of $\lambda$ and remains extremely challenging given present observational capabilities.

In conclusion, while the Big Bang singularity remains unresolved, our work demonstrates that the universe could conceal hidden complexity in the form of evolving wormhole structures supported by thin shells of exotic matter. These configurations challenge our understanding of cosmological homogeneity and suggest that even standard FRW cosmologies may conceal nontrivial topological structures sustained by localized exotic matter.

\begin{acknowledgments}
This work was supported by the Dirección de Investigación y Creación Artística at the Universidad del Bío-Bío through grants No. RE2320220 (MC), GI2310339 (AC, MC and PL) and RE2320212 (PL).
\end{acknowledgments}


\end{document}